\documentclass[]{interact}

\usepackage[numbers,sort&compress]{natbib}
\usepackage{amsmath,amssymb,bm}
\usepackage{graphicx}
\usepackage{booktabs}
\usepackage{siunitx}
\usepackage{xcolor}
\usepackage{hyperref}
\usepackage{lmodern}
\bibpunct[, ]{[}{]}{,}{n}{,}{,}

\title{Electrical transport in ultra-thin films: from Fuchs-Sondheimer to quantum-confinement}

\author{
\name{Alessio Zaccone\textsuperscript{a}\thanks{CONTACT Alessio Zaccone. Email: [alessio.zaccone@unimi.it](mailto:alessio.zaccone@unimi.it)}}
\affil{\textsuperscript{a}Department of Physics ``A. Pontremoli'', University of Milan, via Celoria 16, 20133 Milan, Italy}
}

\begin{document}

\maketitle

\begin{abstract}
Ultra-thin films are fundamental components of modern nanoelectronics, where reducing thickness to the few-nanometer scale leads to a dramatic increase in electrical resistivity. For decades, this behavior has been interpreted in terms of classical size effects, primarily surface scattering within the Fuchs--Sondheimer theory and grain-boundary scattering in the Mayadas--Shatzkes model. While these approaches successfully describe transport when the film thickness is comparable to the electronic mean free path, growing experimental evidence indicates that they become insufficient under extreme confinement. This review discusses the crossover from classical scattering to a quantum-confinement regime in which the electronic states available for transport are fundamentally restructured by finite size. We review the recently proposed reciprocal-space confinement theory, which predicts an exponential increase of resistivity with decreasing thickness at the nanoscale, and discuss how it can be combined with classical surface-scattering models to provide a unified description of ultra-thin metallic and semiconducting films. Finally, we summarize recent experimental evidence supporting this picture and discuss its implications for future nanoelectronic devices, nanoscale interconnects, and quantum transport under extreme spatial confinement.
\end{abstract}


\section{Introduction}

For more than five decades, the extraordinary progress of microelectronics has been driven by the continuous scaling of semiconductor technology, commonly referred to as Moore's law~\cite{Moore1965,Shalf2020}. This evolution, together with the scaling principles established by Dennard \textit{et al.}~\cite{Dennard1974}, enabled an exponential increase in transistor density while simultaneously improving performance and reducing manufacturing costs. As device dimensions have now entered the few-nanometer regime, however, further scaling is increasingly limited by both physical and economic constraints. In particular, the electrical resistance of interconnects, contacts and ultra-thin conducting layers has emerged as one of the principal bottlenecks for future integrated circuits~\cite{HennessyPatterson2019,Davies2025}.

Modern nanoelectronic devices routinely employ metallic and semiconducting films only a few nanometers thick as interconnects, gate electrodes, diffusion barriers and contact layers. As these dimensions approach fundamental electronic length scales such as the electron mean free path and Fermi wavelength, their electrical resistivity increases dramatically, leading to larger RC delays, increased power dissipation and reduced device reliability. Understanding the microscopic origin of this resistivity increase has therefore become a central problem in condensed matter physics as well as a key technological challenge for next-generation nanoelectronics.

The classical description of electrical transport in thin films attributes the increase in resistivity to enhanced electron scattering at external surfaces, interfaces and grain boundaries. The pioneering theories of Fuchs~\cite{Fuchs1938} and Sondheimer~\cite{Sondheimer1952}, later generalized by Mayadas and Shatzkes for polycrystalline materials~\cite{MayadasShatzkes1970}, have provided the theoretical foundation of thin-film transport for more than seventy years. More recently, first-principles electronic-structure calculations have considerably improved the quantitative description of electron-surface scattering, enabling predictive calculations of resistivity in technologically relevant interconnect materials without adjustable parameters~\cite{Yuanyue}.

Although remarkably successful over a broad range of thicknesses, these approaches share a common assumption: the electronic structure itself remains essentially unchanged, while confinement acts only by introducing additional scattering events. As the thickness approaches only a few nanometers, however, this separation between electronic structure and scattering becomes questionable. Spatial confinement progressively suppresses long-wavelength electronic states, reconstructing the available reciprocal-space manifold and modifying the density of transport-active states. Consequently, quantum confinement becomes an intrinsic component of the transport problem rather than a perturbative correction to classical scattering.

The aim of this review is to discuss this new extreme confinement regime and its consequences for electrical transport in ultra-thin metallic and semiconducting films. We focus on the emerging picture in which reciprocal-space confinement provides a unified microscopic framework capable of explaining the rapid increase of resistivity observed experimentally once the film thickness approaches only a few nanometers. In semiconductors, the mechanism manifests itself primarily through confinement-induced carrier depletion, whereas in metals it reduces the transport-active electronic phase space while acting in concert with classical surface scattering. Together, these effects naturally lead to the characteristic exponential dependence
\begin{equation}
\rho(L)\sim\exp\!\left(\frac{C}{\sqrt{L}}\right),
\label{eq:exp-law}
\end{equation}
which has recently found direct experimental support in both nanometric semiconductor films~\cite{Duffy2019,Zaccone2025PRM} and single-crystalline metallic nanofilms~\cite{Yang2025PRL}.

The remainder of this review is organized as follows. Section~2 briefly revisits the classical theories of electron transport in thin metallic films and discusses their range of validity. Section~3 introduces the reciprocal-space confinement theory and derives its consequences for the electronic density of states, carrier concentration and transport coefficients. Section~4 compares the theoretical predictions with recent experimental measurements on semiconducting and metallic ultra-thin films. Finally, the implications of the confinement picture for future nanoelectronic technologies and open theoretical challenges are discussed.

\section{Standard transport theory in thin films}

The theoretical description of electrical conduction in solids originates from the classical Drude model and its quantum extension due to Sommerfeld, which remain the foundation of electron transport theory in metals and degenerate semiconductors~\cite{MottJones1936,Kittel,AshcroftMermin,Ziman1960}. Within this framework, electrical conduction arises from the motion of free carriers undergoing scattering by phonons, impurities and structural defects. The conductivity is given by the familiar Drude--Sommerfeld expression
\begin{equation}
\sigma = ne\mu = \frac{ne^2\tau}{m^\ast},
\label{eq:drude}
\end{equation}
where $n$ is the free-carrier concentration, $e$ is the elementary charge, $\mu$ is the carrier mobility, $\tau$ is the relaxation time and $m^\ast$ is the effective carrier mass. The corresponding resistivity is simply
\begin{equation}
\rho=\sigma^{-1}.
\end{equation}

For bulk conductors, the carrier concentration is determined by the electronic structure of the material, whereas the mobility reflects the various scattering mechanisms experienced by the carriers. The central assumption underlying classical transport theory is that these two quantities may be treated independently: the electronic structure determines the number of carriers, while scattering determines how efficiently they conduct.

When one dimension of the conductor becomes comparable to the electronic mean free path, additional scattering at external surfaces and interfaces reduces the carrier mobility, giving rise to the classical size effect. This picture forms the basis of the Fuchs--Sondheimer theory for single-crystalline films~\cite{Fuchs1938,Sondheimer1952} and of the Mayadas--Shatzkes model for polycrystalline conductors~\cite{MayadasShatzkes1970}. More recently, first-principles electronic-structure calculations have considerably refined this description by treating electron--surface scattering atomistically and enabling predictive calculations of resistivity without adjustable parameters~\cite{Yuanyue}.

The common feature of all these approaches is that confinement influences transport only indirectly through scattering, while the electronic structure and carrier concentration remain essentially unchanged. As will be discussed in the following sections, this assumption eventually breaks down once the film thickness approaches only a few nanometers, where quantum confinement modifies the available electronic states themselves.

\begin{figure}[h]
\centering
\includegraphics[width=0.93\linewidth]{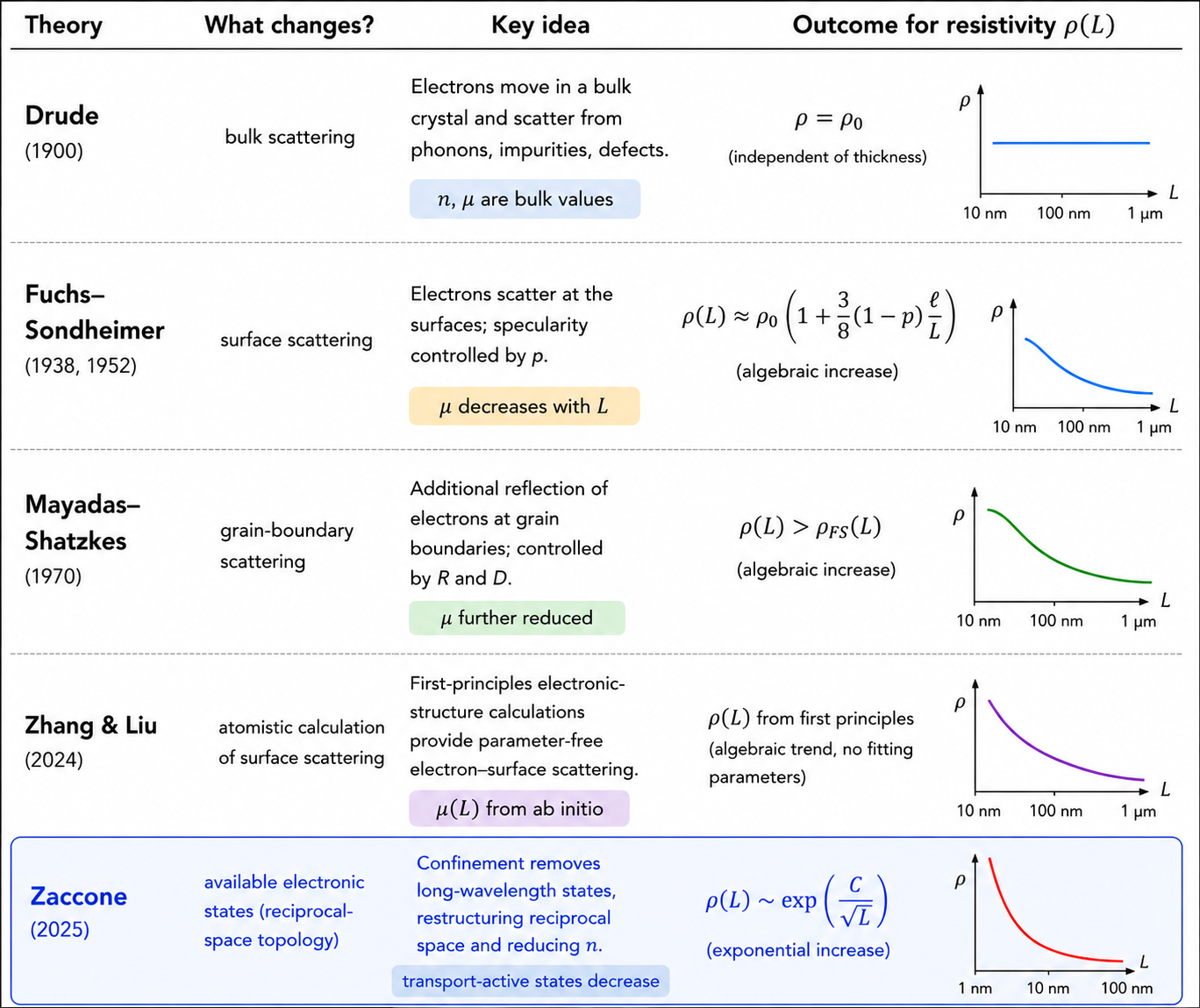}
\caption{Evolution of theoretical descriptions of electrical transport in thin films. Classical transport begins with the Drude theory, followed by boundary scattering in the Fuchs–Sondheimer model and grain-boundary scattering in the Mayadas–Shatzkes theory. Recent first-principles calculations by Zhang and Liu \cite{Yuanyue} compute surface scattering from electronic structure without phenomenological parameters. The quantum-confinement theory developed by Zaccone \cite{Zaccone2025PRM} introduces a complementary mechanism in which confinement restructures the available electronic states in reciprocal space, producing carrier depletion and an exponential increase of resistivity upon decreasing the thickness at the nanoscale. }
\label{fig1}
\end{figure}

\begin{figure}[h]
\centering
\includegraphics[width=0.93\linewidth]{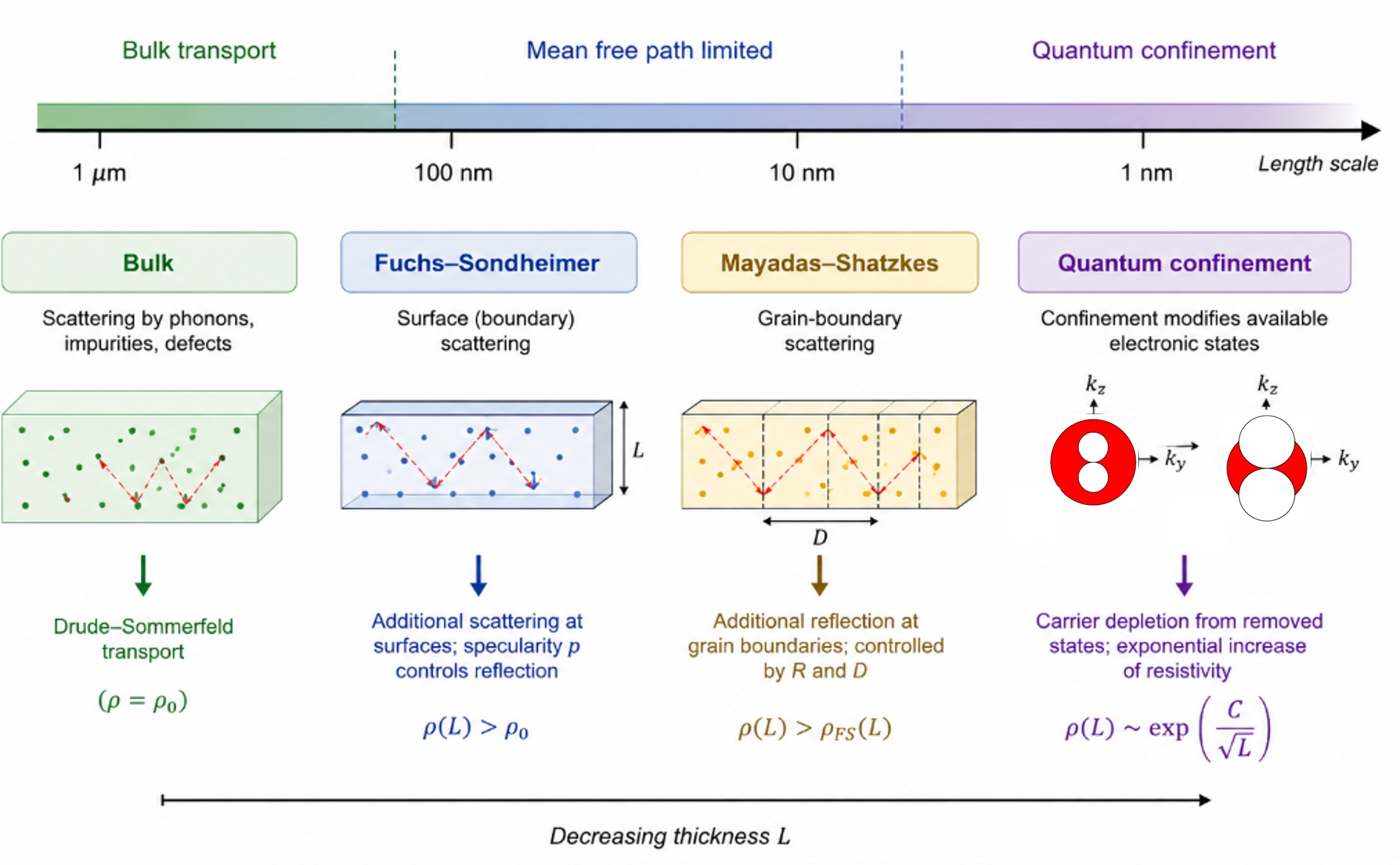}
\caption{Hierarchy of transport mechanisms governing the electrical resistivity of ultra-thin films as the characteristic dimension is progressively reduced.
In bulk conductors ($L\gg\ell$), electron transport is governed by intrinsic scattering from phonons, impurities and defects, and is well described by the Drude--Sommerfeld model. When the film thickness $L$ becomes comparable to the electron mean free path $\ell$, additional boundary scattering gives rise to the classical size effect described by the Fuchs--Sondheimer theory, while grain-boundary scattering further increases the resistivity in polycrystalline films according to the Mayadas--Shatzkes model. As confinement approaches only a few nanometers, quantum effects become dominant: the available electronic states in reciprocal space are modified by spatial confinement, leading to a reduction of the carrier concentration and an exponential increase of the resistivity,
$\rho(L)\sim\exp\!\left(\frac{C}{\sqrt{L}}\right).$
The figure emphasizes the successive physical mechanisms that become relevant as transport evolves from the bulk regime toward the quantum-confinement limit.
}
\label{fig2}
\end{figure}

\section{Classical size effects in thin-film conduction}

\subsection{Mean free path and geometrical limitation}

The electron mean free path $\ell$ is the average distance travelled by a conduction electron between momentum-relaxing scattering events. The concept of the electronic mean free path has played a central role in transport theory since the early development of the semiclassical Boltzmann description of metals \cite{Ziman1960}. In a bulk metal, $\ell$ is controlled by phonons, impurities, defects and electron-electron processes. In a film of thickness $L$, an additional length scale enters the problem. When $L \gg \ell$, the boundaries are rarely sampled and the material behaves approximately as a bulk conductor. When $L \sim \ell$, surface scattering becomes important and the resistivity increases.

The classical conductivity can be expressed as
\begin{equation}
\sigma_0 = \frac{n e^2 \ell}{m^\ast v_F},
\label{eq:sigma_mfp}
\end{equation}
where $v_F$ is the Fermi velocity. Equivalently,
\begin{equation}
\rho_0 = \frac{m^\ast v_F}{n e^2 \ell}.
\end{equation}
In the classical size-effect framework, the main role of finite thickness is to replace the bulk mean free path $\ell$ by an effective thickness-dependent mean free path $\ell_{\rm eff}(L)$, while $n$ is kept fixed:
\begin{equation}
\rho(L) = \frac{m^\ast v_F}{n e^2 \ell_{\rm eff}(L)}.
\end{equation}

\subsection{Fuchs--Sondheimer surface scattering}

The Fuchs--Sondheimer theory treats electron transport in a thin metallic film by solving the Boltzmann transport equation with boundary conditions at the two film surfaces. Originally proposed independently by K. Fuchs and later developed in detail by Sondheimer, this theory represents one of the cornerstones of modern thin-film transport \cite{Fuchs1938,Sondheimer1952}. The essential parameter is the specularity coefficient $p$, where $p=1$ corresponds to perfectly specular reflection and $p=0$ to completely diffuse scattering.

For a film of thickness $L$, the Fuchs--Sondheimer result may be written as
\begin{equation}
\frac{\rho_{\rm FS}(L)}{\rho_0}
=
\left[
1-\frac{3}{2\kappa}(1-p)
\int_{1}^{\infty}
\left(
\frac{1}{t^3}-\frac{1}{t^5}
\right)
\frac{1-\exp(-\kappa t)}
{1-p\exp(-\kappa t)}
dt
\right]^{-1},
\label{eq:FS}
\end{equation}
where
\begin{equation}
\kappa = \frac{L}{\ell}.
\end{equation}
Equation (\ref{eq:FS}) has been successfully applied to numerous metallic systems including Cu, Al and Ag thin films \cite{Zhang2004}.
In the thick-film limit $L\gg \ell$, Eq.~\eqref{eq:FS} reduces to
\begin{equation}
\frac{\rho_{\rm FS}(L)}{\rho_0}
\simeq
1+\frac{3}{8}(1-p)\frac{\ell}{L}.
\label{eq:FS_asymptotic}
\end{equation}
Thus, the classical surface-scattering correction is approximately algebraic in $1/L$, not exponential.

Although the Fuchs--Sondheimer theory has remained the standard framework for describing classical size effects for more than seven decades \cite{Fuchs1938,Sondheimer1952}, recent first-principles calculations have shown that electron-surface scattering can now be computed directly without introducing phenomenological specularity parameters \cite{Yuanyue,ZhouGall2018}.

\subsection{Mayadas--Shatzkes grain-boundary scattering}

For polycrystalline films, grain boundaries provide an additional source of momentum relaxation. The Mayadas--Shatzkes model extends the classical picture by introducing grain-boundary reflection as an additional scattering mechanism \cite{MayadasShatzkes1970}. The Mayadas--Shatzkes model describes grain boundaries as partially reflecting planes. For specular surface scattering and grain-boundary reflection coefficient $R$, the resistivity enhancement can be written as
\begin{equation}
\frac{\rho_{\rm MS}}{\rho_0}
=
\left[
1-\frac{3}{2}\alpha
+3\alpha^2
-3\alpha^3\ln\left(1+\frac{1}{\alpha}\right)
\right]^{-1},
\label{eq:MS}
\end{equation}
with
\begin{equation}
\alpha = \frac{\ell}{D}\frac{R}{1-R},
\label{eq:alpha_MS}
\end{equation}
where $D$ is the average grain size.

A useful practical approximation is to regard the total resistivity enhancement as arising from both surface and grain-boundary scattering:
\begin{equation}
\rho_{\rm cl}(L) \approx \rho_{\rm FS}(L) + \rho_{\rm MS}(L) - \rho_0.
\label{eq:classical_combined}
\end{equation}
This expression captures the classical size-effect picture: resistivity increases because the mobility decreases.

\section{Limitations of the classical picture}

The Fuchs--Sondheimer and Mayadas--Shatzkes theories explain a broad range of thin-film transport data at thicknesses comparable to, but not drastically smaller than, the electron mean free path. Numerous experimental studies have demonstrated that these classical models provide excellent agreement for thicknesses comparable to the electron mean free path but progressively underestimate the resistivity of films below approximately $10$ nm \cite{Zhang2004,Duffy2019}. They are especially useful for interconnect metals such as Cu, Al and Ag, where the electron mean free path at room temperature is of order tens of nanometers. However, the few-nanometer regime raises three difficulties.

First, the observed resistivity increase can be much steeper than the algebraic dependence expected from Eq.~\eqref{eq:FS_asymptotic}. Second, fitting the data often requires unphysical values of the surface specularity or grain-boundary reflection coefficient. Third, Hall measurements on ultrathin Al films directly revealed a reduction of the carrier concentration with decreasing thickness \cite{Du2004}. This directly violates the assumption that $n$ remains bulk-like while only the mobility changes.

The conductivity should therefore be written more generally as
\begin{equation}
\sigma(L) = n(L)e\mu(L),
\label{eq:sigma_general}
\end{equation}
where both $n(L)$ and $\mu(L)$ can depend on thickness. Classical size-effect theories mainly address $\mu(L)$. Quantum confinement modifies $n(L)$.

\subsection{Beyond phenomenological surface scattering}
Although the Fuchs--Sondheimer theory has been extraordinarily successful,
its central quantity, the specularity coefficient $p$, remains a
phenomenological parameter whose value must generally be inferred from
experiment. Likewise, the grain-boundary reflection coefficient $R$ in the
Mayadas--Shatzkes theory is not determined microscopically.

Recent progress in first-principles electronic-structure calculations has
made it possible to compute electron-surface scattering without introducing
empirical fitting parameters. In particular, Zhang and Liu developed a
parameter-free first-principles framework in which electron-surface
scattering is obtained directly from the electronic structure of the
confined material, eliminating the need for an assumed surface specularity
parameter \cite{Yuanyue}. Their calculations reveal that the
surface-scattering strength depends strongly on the crystallographic
orientation and the detailed electronic structure of the surface, providing
a microscopic foundation for transport in ultrathin metallic films. 

These first-principles approaches represent a major conceptual advance over classical size-effect theories because they remove phenomenological parameters from the description of boundary scattering. Nevertheless, they retain the implicit assumption that the available electronic states are those of the confined material, and therefore describe only the scattering among these states. In the next section we discuss a complementary quantum
mechanism in which sufficiently strong confinement modifies the available electronic states themselves through a restructuring of reciprocal space, leading to a thickness-dependent carrier population and ultimately to an exponential increase of the resistivity \cite{Zaccone2025PRM}.

\section{Quantum confinement and suppression of electronic states}
Recent theoretical work demonstrated that this behavior arises naturally from a confinement-induced restructuring of the electronic states in reciprocal space, leading to a topological change of the available momentum-space manifold \cite{Zaccone2025PRM}.

\begin{figure*}[t]
\centering
\includegraphics[width=\textwidth]{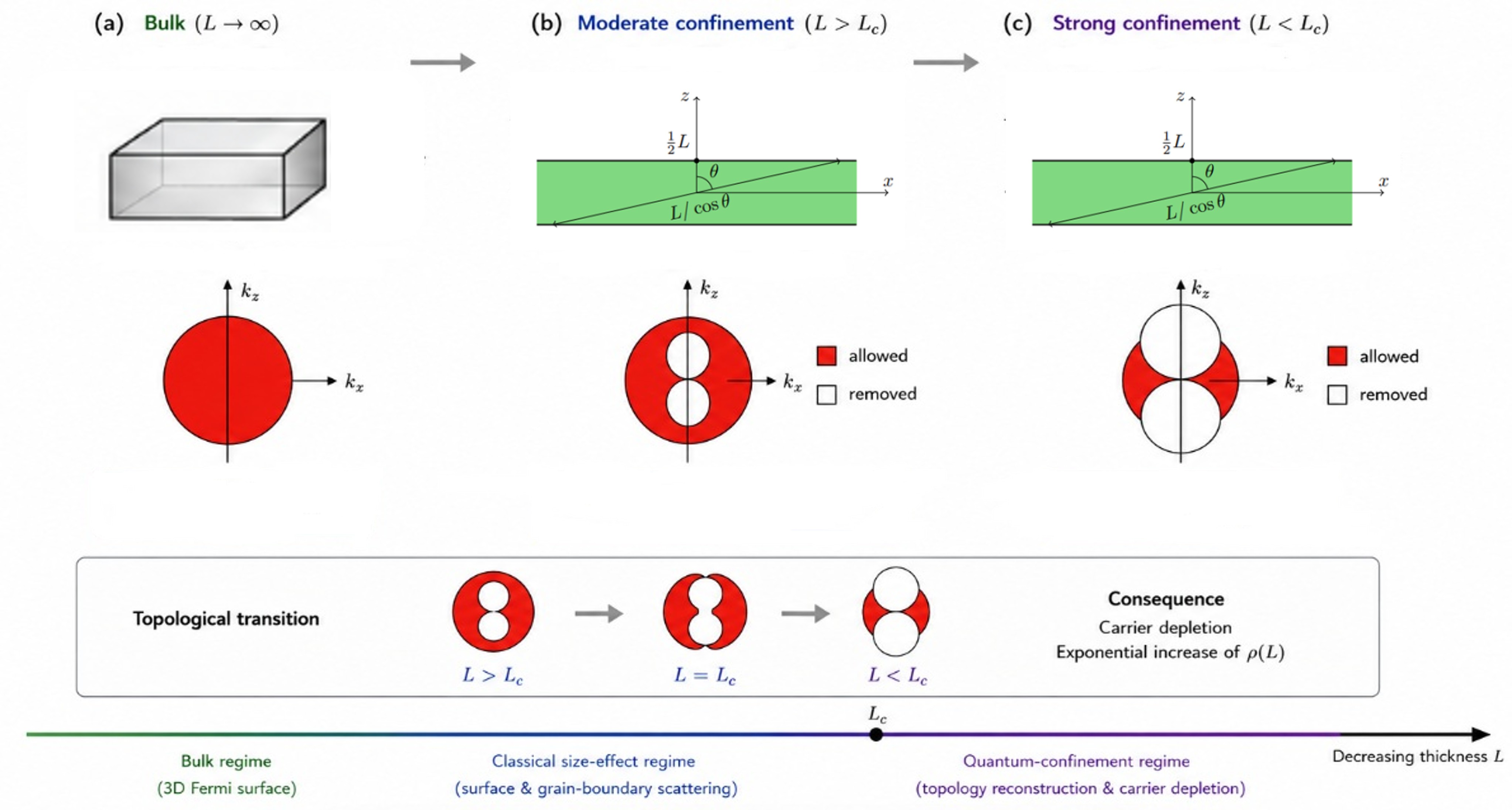}
\caption{Evolution of the reciprocal-space topology under thickness confinement. (a) In the bulk limit ($L\rightarrow\infty$), all electronic momentum states within the Fermi sphere are available, giving rise to a simply connected reciprocal-space manifold. (b) As the thickness decreases, confinement progressively suppresses long-wavelength states along the confinement direction, removing portions of reciprocal space while preserving the overall connectivity of the available states. (c) Below the critical thickness $L_c$, the excluded regions merge, producing a topological transition in the accessible reciprocal-space manifold. The remaining allowed states split into disconnected electron pockets separated by confinement-forbidden regions. The lower schematic illustrates the evolution of the topology across the critical thickness $L_c$, marking the crossover from the classical size-effect regime to the quantum-confinement regime. This confinement-induced reconstruction of reciprocal space reduces the number of transport-active carriers and provides the microscopic origin of the exponential scaling of the resistivity, $\rho(L)\sim\exp(C/\sqrt{L})$.}
\label{fig:topology}
\end{figure*}

The physical origin of the confinement-induced transport anomaly is illustrated schematically in Fig.~\ref{fig:topology}. As the film thickness decreases, long-wavelength electronic states become progressively inaccessible, producing an evolution of the topology of the available reciprocal-space manifold. Above the critical thickness $L_c$ the available states remain simply connected, whereas below $L_c$ the confinement-forbidden regions merge, leading to a topological transition in reciprocal space. This reconstruction of the electronic-state manifold reduces the number of transport-active carriers and ultimately gives rise to the exponential increase in resistivity discussed below.

Consider a film confined along the (arbitrarily chosen) $z$-direction and extended in the $xy$-plane. In the bulk, conduction electrons occupy states within a Fermi sphere in reciprocal space. In a film of thickness $L$, electronic wavelengths exceeding the confinement length along a given propagation direction cannot be accommodated. This removes portions of the available momentum-space volume.

For a carrier propagating at polar angle $\theta$ with respect to the confinement direction, the maximum wavelength compatible with the film thickness is
\begin{equation}
\lambda_{\max} = \frac{L}{\cos\theta}.
\end{equation}
Equivalently, long-wavelength states are suppressed in reciprocal space. The resulting state depletion can be represented as the removal of hole-pocket-like regions from the original Fermi sphere.

At moderate confinement, the available reciprocal-space volume may be written schematically as
\begin{equation}
V_k(L)
=
\frac{4\pi}{3}k_F^3
-
2\frac{4\pi}{3}
\left(\frac{\pi}{L}\right)^3,
\label{eq:Vk_moderate}
\end{equation}
where $k_F$ is the bulk Fermi wave vector. Below a critical thickness, the suppressed regions intersect strongly with the Fermi surface, producing a topological reconstruction of the available electronic states. The critical scale is
\begin{equation}
L_c = \left(\frac{2\pi}{n}\right)^{1/3}.
\label{eq:Lc}
\end{equation}

In the strongly confined regime, the available reciprocal-space volume scales differently and the Fermi level becomes thickness dependent, according to a mathematical model first proposed for free-electrons confined in thin films in Ref. \cite{Travaglino_2023} (similar confinement ideas have been applied also to Bose-Einstein condensates \cite{Travaglino_2022} and to phonons \cite{Heat,thermal}).

The confinement-induced reconstruction of reciprocal space also modifies the electronic density of states. For free carriers with parabolic dispersion,
\begin{equation}
    \epsilon=\frac{\hbar^2 k^2}{2m},
    \qquad
    k=\frac{\sqrt{2m\epsilon}}{\hbar}.
\end{equation}
The crossover between the bulk-like and strongly confined regimes occurs at
\begin{equation}
    k^\ast=\frac{2\pi}{L},
    \qquad
    \epsilon^\ast
    =
    \frac{\hbar^2(k^\ast)^2}{2m}
    =
    \frac{2\pi^2\hbar^2}{mL^2}.
\end{equation}
For $\epsilon>\epsilon^\ast$, the subtraction of the two confinement-forbidden spheres is independent of $k$, and the usual three-dimensional free-electron density of states is recovered:
\begin{equation}
    g(\epsilon)
    =
    \frac{V(2m)^{3/2}}{2\pi^2\hbar^3}\epsilon^{1/2}.
\end{equation}
For $\epsilon<\epsilon^\ast$, the available reciprocal-space volume is instead
\begin{equation}
    {\rm Vol}_k=\frac{Lk^4}{2},
\end{equation}
so that
\begin{equation}
    N(\epsilon_0,\epsilon)
    =
    \frac{V}{(2\pi)^3}{\rm Vol}_k
    =
    \frac{V}{(2\pi)^3}\frac{Lk^4}{2}.
\end{equation}
Using $k^4=(2m)^2\epsilon^2/\hbar^4$ and differentiating with respect to $\epsilon$ gives
\begin{equation}
    g(\epsilon)
    =
    \frac{VLm^2}{2\pi^3\hbar^4}\epsilon,
    \qquad
    \epsilon<\epsilon^\ast .
\end{equation}
Thus the density of states becomes
\begin{equation}
g(\epsilon)=
\begin{cases}
\dfrac{VLm^2}{2\pi^3\hbar^4}\epsilon,
& \epsilon<\dfrac{2\pi^2\hbar^2}{mL^2}, \\[1.2em]
\dfrac{V(2m)^{3/2}}{2\pi^2\hbar^3}\epsilon^{1/2},
& \epsilon>\dfrac{2\pi^2\hbar^2}{mL^2}.
\end{cases}
\label{eq:confined_DOS}
\end{equation}
The important consequence is that the low-energy DOS changes from the usual three-dimensional form $g(\epsilon)\propto \epsilon^{1/2}$ to the confinement-controlled form $g(\epsilon)\propto \epsilon$.

In the strongly confined regime, $L<L_c$, the Fermi energy lies below the crossover energy, $\epsilon_F<\epsilon^\ast$, and the occupied states are counted using the linear DOS. At zero temperature,
\begin{equation}
    N
    =
    \int_0^{\epsilon_F} g_s g(\epsilon)\,d\epsilon,
\end{equation}
with spin degeneracy $g_s=2$. Hence,
\begin{equation}
    N
    =
    \int_0^{\epsilon_F}
    2\frac{VLm^2}{2\pi^3\hbar^4}\epsilon\,d\epsilon
    =
    \frac{VLm^2}{2\pi^3\hbar^4}\epsilon_F^2.
\end{equation}
Equivalently,
\begin{equation}
    N
    =
    \frac{VL(2m)^2}{(2\pi)^3\hbar^4}\epsilon_F^2.
    \label{eq:N_confined}
\end{equation}
Writing the carrier density as $n=N/V$, Eq.~\eqref{eq:N_confined} gives
\begin{equation}
    n
    =
    \frac{L(2m)^2}{(2\pi)^3\hbar^4}\epsilon_F^2,
\end{equation}
and therefore
\begin{equation}
    \epsilon_F
    =
    \frac{\hbar^2}{m}
    \left[
    \frac{(2\pi)^3 n}{L}
    \right]^{1/2}.
\end{equation}
At zero temperature the chemical potential is equal to the Fermi energy, $\mu=\epsilon_F$, so that in the strongly confined regime
\begin{equation}
    \mu(L)
    =
    \frac{\hbar^2}{m}
    \left[
    \frac{(2\pi)^3 n}{L}
    \right]^{1/2},
    \qquad L<L_c.
    \label{eq:mu_confined}
\end{equation}
This result shows explicitly that confinement drives an upward shift of the chemical potential as $\mu(L)\propto L^{-1/2}$.

This differs fundamentally from the bulk expression because the topology and geometry of the accessible momentum-space manifold have changed.

The confinement-induced increase of the chemical potential described by Eq.~\eqref{eq:mu_confined} has different physical consequences depending on the electronic character of the material.

For intrinsic and weakly doped semiconductors, the free-carrier concentration is determined by thermal excitation across the energy gap. An increase of the chemical potential therefore corresponds to an effective widening of the activation energy for carrier generation, leading to a reduction of the equilibrium carrier concentration. In this case, quantum confinement acts primarily by modifying the number of transport-active carriers.

In good metals, by contrast, the total electron concentration is essentially fixed by the valence electron density and is only weakly affected by confinement. Nevertheless, the reconstruction of reciprocal space derived above still modifies the manifold of electronic states available for transport. While the total number of electrons remains nearly constant, the number of states participating efficiently in electrical conduction is reduced because long-wavelength states are progressively eliminated by confinement. Consequently, the conductivity can decrease even in the absence of a significant change in the total carrier density.

These two limiting cases share the same microscopic origin, namely the confinement-induced reconstruction of the available reciprocal-space volume, but differ in how this reconstruction affects electrical transport.

In full generality, the conductivity may therefore be written as
\begin{equation}
\sigma(L)=e\,n(L)\,\mu(L),
\label{eq:sigma_general_review}
\end{equation}
where both the carrier concentration and the mobility may become thickness dependent.

For semiconductors, the dominant confinement effect is the reduction of the free-carrier concentration,
\begin{equation}
n(L)<n_0,
\end{equation}
whereas the mobility is mainly controlled by conventional scattering processes,
\begin{equation}
\mu(L)\simeq\mu_{\rm FS}(L),
\end{equation}
with $\mu_{\rm FS}(L)$ given by the Fuchs--Sondheimer description together with possible grain-boundary corrections.

For metals, on the other hand, the carrier concentration remains approximately constant,
\begin{equation}
n(L)\simeq n_0,
\end{equation}
while confinement modifies the transport efficiency of the electronic states through the reduction of the available reciprocal-space manifold. This effect may be represented phenomenologically through a confinement factor $\eta_{\rm QC}(L)$,
\begin{equation}
\mu(L)=\mu_{\rm FS}(L)\,\eta_{\rm QC}(L),
\end{equation}
where $\eta_{\rm QC}(L)\le1$ accounts for the reduction of the transport-active phase space induced by quantum confinement.

The remainder of this section focuses first on intrinsic and weakly doped semiconductors, for which the confinement-induced carrier depletion can be derived analytically and leads directly to an exponential thickness dependence of the resistivity. Subsequently, we shall show that the same reciprocal-space mechanism also provides a natural framework for understanding the recently observed exponential increase of resistivity in ultra-thin single-crystalline metallic films.

\section{Confinement-induced carrier depletion in semiconductors}
We first consider intrinsic and weakly doped semiconductors, for which the free-carrier concentration is determined by thermal excitation across the band gap. Because the chemical potential increases according to Eq.~\eqref{eq:mu_confined}, confinement produces an effective widening of the activation energy and hence a reduction of the equilibrium carrier concentration. This mechanism leads naturally to an exponential increase of the resistivity as the film thickness is reduced.
For intrinsic or weakly doped semiconductors, the carrier concentration is controlled by the energy separation between the Fermi level and the relevant band edge. If confinement increases the effective energy gap or shifts the chemical potential, the carrier concentration is exponentially suppressed. The intrinsic carrier concentration may be written as
\begin{equation}
n_i(L)
=
\sqrt{N_c(L)N_v(L)}
\exp\left[-\frac{E_g(L)}{2k_B T}\right],
\label{eq:ni}
\end{equation}
where $N_c$ and $N_v$ are the effective densities of states in the conduction and valence bands, respectively, and $E_g(L)$ is the confinement-dependent effective gap.

If the confinement-induced energy shift scales as
\begin{equation}
\Delta E(L) \sim \frac{A}{\sqrt{L}},
\label{eq:DeltaE}
\end{equation}
then the carrier concentration decreases as
\begin{equation}
n(L) \sim n_0
\exp\left[-\frac{A}{2k_B T\sqrt{L}}\right].
\label{eq:n_exp}
\end{equation}
Assuming that the mobility varies more slowly than this exponential factor, the resistivity becomes
\begin{equation}
\rho(L)
=
\frac{1}{e n(L)\mu(L)}
\sim
\rho_0
\exp\left(\frac{C}{\sqrt{L}}\right),
\label{eq:rho_exp}
\end{equation}
where
\begin{equation}
C = \frac{A}{2k_B T}.
\end{equation}
This exponential scaling was first derived theoretically in Ref.~\cite{Zaccone2025PRM}.

Equation~\eqref{eq:rho_exp} is the central quantum-confinement result. It predicts that the resistivity does not merely increase as a power of inverse thickness but instead grows exponentially as the film enters the few-nanometer regime.

\section{Combining Fuchs--Sondheimer theory with quantum confinement}

The preceding sections have shown that classical surface scattering and quantum confinement affect electrical transport through distinct microscopic mechanisms. Surface scattering reduces the carrier mobility without significantly altering the electronic structure, whereas quantum confinement modifies the set of electronic states available for conduction. These two effects therefore enter naturally as independent factors in the conductivity.

For intrinsic and weakly doped semiconductors, the conductivity may be written as
\begin{equation}
\sigma(L)
=
n(L)e\mu_{\rm FS}(L),
\end{equation}
where the mobility $\mu_{\rm FS}(L)$ is described by the classical Fuchs--Sondheimer theory and the carrier concentration $n(L)$ follows from the confinement theory discussed in the previous sections. Introducing
\begin{equation}
\mu_{\rm FS}(L)=\mu_0f_{\rm FS}(L),
\end{equation}
and
\begin{equation}
n(L)=n_0f_{\rm QC}(L),
\end{equation}
one obtains
\begin{equation}
\sigma(L)
=
n_0e\mu_0
f_{\rm FS}(L)
f_{\rm QC}(L),
\end{equation}
or, equivalently,
\begin{equation}
\rho(L)
=
\rho_{\rm FS}(L)
\frac{1}{f_{\rm QC}(L)}.
\label{eq:combined_general}
\end{equation}

Using the confinement law derived in Ref.~\cite{Zaccone2025PRM},
\begin{equation}
f_{\rm QC}(L)
=
\exp\!\left(-\frac{C}{\sqrt{L}}\right),
\end{equation}
one immediately obtains
\begin{equation}
\rho(L)
=
\rho_{\rm FS}(L)
\exp\!\left(\frac{C}{\sqrt{L}}\right),
\label{eq:FS_Zaccone}
\end{equation}
which provides a unified description of classical surface scattering and quantum confinement for ultra-thin semiconductor films. At large thicknesses the exponential factor approaches unity and the classical Fuchs--Sondheimer limit is recovered, whereas below a few nanometers the confinement contribution dominates the transport.

\subsection{Extension to metallic thin films}

The analytical derivation presented above applies rigorously to intrinsic and weakly doped semiconductors, where quantum confinement modifies the free-carrier concentration through the confinement-induced shift of the chemical potential. Good metals require a different interpretation. Since the conduction-electron density is primarily fixed by the valence electron concentration, the total number of electrons is not expected to change appreciably with thickness. Nevertheless, the reciprocal-space reconstruction induced by quantum confinement remains exactly the same.

Instead of depleting the total carrier concentration, confinement is therefore expected to reduce the fraction of electronic states that effectively participate in transport. This observation naturally suggests writing the conductivity of metallic thin films as
\begin{equation}
\sigma(L)
=
e\,n_0\,
\mu_{\rm FS}(L)\,
\eta_{\rm QC}(L),
\label{eq:master_transport}
\end{equation}
where $n_0$ is the bulk conduction-electron density, $\mu_{\rm FS}(L)$ is the mobility limited by classical surface scattering, and $\eta_{\rm QC}(L)\le1$ is a confinement factor representing the reduction of the transport-active reciprocal-space manifold.

The microscopic derivation of $\eta_{\rm QC}(L)$ directly from the confined electronic structure remains an important open theoretical problem. Nevertheless, the reciprocal-space reconstruction developed in the previous sections naturally suggests that $\eta_{\rm QC}(L)$ should decrease continuously as confinement progressively removes long-wavelength electronic states. Motivated by this physical picture, and by the recent measurements on single-crystalline Au films by Yang \textit{et al.}~\cite{Yang2025PRL}, we propose the asymptotic form
\begin{equation}
\eta_{\rm QC}(L)
\sim
\exp\!\left(-\frac{C}{\sqrt{L}}\right),
\end{equation}
which immediately gives
\begin{equation}
\rho(L)
=
\rho_{\rm FS}(L)
\exp\!\left(\frac{C}{\sqrt{L}}\right).
\label{eq:FS_QC}
\end{equation}

Equation~\eqref{eq:FS_QC} should therefore be viewed as a physically motivated extension of the semiconductor confinement theory to metallic films. While the exponential confinement factor follows analytically for weakly doped semiconductors, its application to metals is presently motivated by the reciprocal-space reconstruction discussed above and is strongly supported by recent experiments. In particular, the transport measurements on single-crystalline Au nanofilms by Yang \textit{et al.}~\cite{Yang2025PRL}, where grain-boundary scattering is essentially absent, provide compelling evidence that reciprocal-space confinement becomes a dominant transport mechanism once metallic films are reduced to only a few nanometers in thickness.

\section{Experimental evidence}

The quantum-confinement theory reviewed in the previous sections predicts that, once the film thickness approaches only a few nanometers, electrical transport is no longer governed solely by classical surface scattering but also by a confinement-induced restructuring of the available electronic states in reciprocal space. Depending on the electronic character of the material, this restructuring manifests itself either as carrier depletion (in semiconductors) or as a reduction of the transport-active reciprocal-space manifold (in metals). In both cases, however, the theory predicts the same characteristic exponential dependence of the resistivity,
\begin{equation}
\rho(L)\sim\exp\!\left(\frac{C}{\sqrt{L}}\right),
\end{equation}
which provides a distinctive experimental signature of the confinement mechanism.

\subsection{Nanometric semiconductor films}

The first quantitative validation of the theory was obtained for nanometric semiconductor films. In Ref.~\cite{Zaccone2025PRM}, the confinement model was shown to reproduce the thickness dependence of the electrical conductivity measured in ultrathin Si films by Duffy \textit{et al.}~\cite{Duffy2019}. In these systems, the confinement-induced increase of the chemical potential leads to a reduction of the free-carrier concentration, giving rise to an exponential increase of the resistivity that cannot be explained within the classical Fuchs--Sondheimer framework alone.

The agreement between theory and experiment demonstrated that carrier depletion induced by quantum confinement becomes the dominant transport mechanism once the film thickness falls below approximately \SI{10}{nm}. This provided the first experimental evidence supporting the reciprocal-space confinement picture developed in Ref.~\cite{Zaccone2025PRM}.

\subsection{Single-crystalline Au at the few-nanometer scale}

A decisive experimental confirmation of the confinement picture has recently been provided by the transport measurements of Yang \textit{et al.}~\cite{Yang2025PRL} on single-crystalline Au nanofilms. Because these epitaxial films are essentially free of grain boundaries, they constitute an ideal platform for isolating the effects of quantum confinement from additional sources of disorder.

The measurements therefore provide strong support for extending the reciprocal-space confinement picture to metallic conductors. While the exponential law follows analytically for weakly doped semiconductors, the agreement obtained for Au strongly suggests that an analogous confinement factor also governs transport in good metals. This increase is substantially stronger than predicted by the classical Fuchs--Sondheimer theory, whereas the combined description developed in the present review,
\begin{equation}
\rho(L)=\rho_{\rm FS}(L)\exp\!\left(\frac{C}{\sqrt{L}}\right),
\end{equation}
accurately reproduces the measurements over the entire confinement range.

The significance of this agreement extends beyond the specific Au system. Unlike the semiconductor case, where confinement primarily modifies the carrier concentration, the Au measurements demonstrate that the same reciprocal-space reconstruction also governs transport in good metals, where the total electron density remains approximately constant but the transport-active phase space is progressively reduced.

\begin{figure*}[t]
\centering
\includegraphics[width=\textwidth]{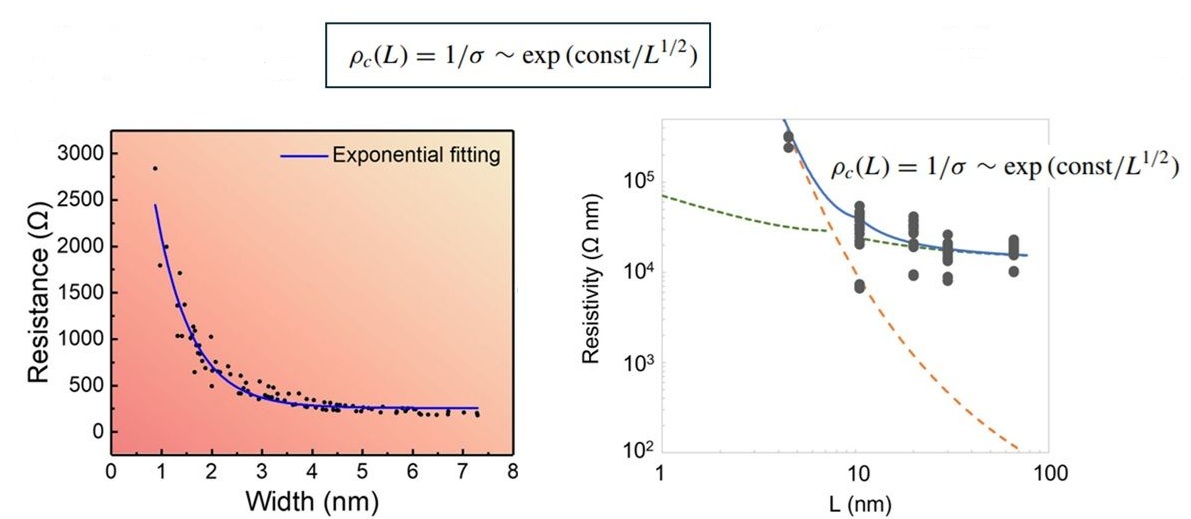}
\caption{Experimental evidence for the exponential quantum-confinement law in semiconductor and metallic ultra-thin films.
Left: Electrical resistivity of single-crystalline Au nanofilms measured by Yang \textit{et al.}~\cite{Yang2025PRL}. The solid blue curve represents the combined Fuchs--Sondheimer plus quantum-confinement model,
$\rho(L)=\rho_{\rm FS}(L)\exp\!\left(\frac{C}{\sqrt{L}}\right),$
whereas the dashed orange curve shows the prediction of the classical Fuchs--Sondheimer theory alone. The classical theory systematically underestimates the resistivity in the few-nanometer regime, while the combined model accurately reproduces the measurements. The agreement obtained for both semiconductor and metallic systems demonstrates that reciprocal-space confinement provides a unified microscopic mechanism for the exponential increase of resistivity in ultra-thin films. Right: Electrical resistance of nanometric Si films as a function of film width (adapted from Refs.~\cite{Duffy2019,Zaccone2025PRM}). The continuous curve is the prediction of the quantum-confinement theory, which accounts for the confinement-induced reduction of the free-carrier concentration and reproduces the rapid increase of resistance below a few nanometers.}
\label{fig:experimental}
\end{figure*}

The experimental evidence accumulated so far therefore reveals a remarkable degree of universality across different classes of materials. Although the microscopic manifestation of quantum confinement differs between semiconductors and metals, the same confinement-induced scaling emerges in both cases once the characteristic dimension approaches only a few nanometers. Figure~\ref{fig:experimental} compares these two representative examples. Together, they strongly suggest that the exponential dependence
\begin{equation}
\rho(L)\sim\exp\!\left(\frac{C}{\sqrt{L}}\right)
\end{equation}
is not merely a material-specific fitting relation, but rather the experimental signature of a universal quantum-confinement mechanism governing electrical transport in ultra-thin films.

\begin{figure}[t]
\centering
\includegraphics[width=0.78\linewidth]{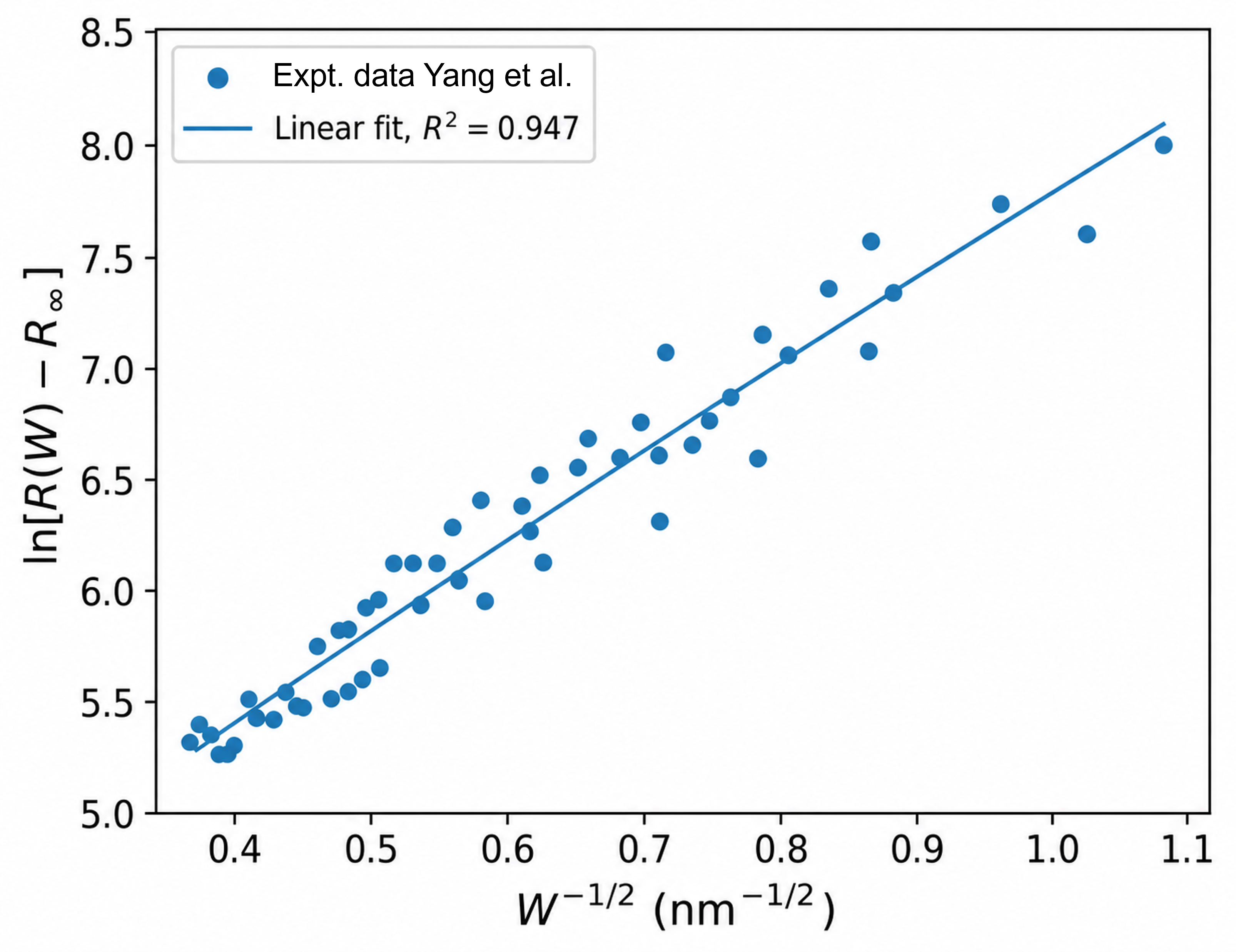}
\caption{Linearized representation of the confinement law for the experimental data of Yang \textit{et al.}~\cite{Yang2025PRL}.
The experimental resistance data were digitized from Ref.~\cite{Yang2025PRL} and replotted according to the Zaccone law prediction \cite{Zaccone2025PRM},
$\ln[R(W)-R_\infty]$ versus $W^{-1/2}$.
The approximately linear dependence ($R^2=0.947$) is consistent with the exponential confinement law
$R(W)-R_\infty\propto\exp(C/\sqrt{W})$ first derived in \cite{Zaccone2025PRM}.
For comparison, a conventional exponential dependence on width,
$R(W)-R_\infty\propto\exp(-BW)$,
gives a lower quality fit ($R^2=0.927$) for the same digitized dataset.
}
\label{fig:linearized}
\end{figure}

To further assess the functional form predicted by the reciprocal-space confinement theory, we reanalyzed the experimental data of Yang \textit{et al.}~\cite{Yang2025PRL} by exploiting the linearized form of Eq.~\eqref{eq:FS_QC}. Writing
\begin{equation}
R(W)=R_\infty+A\exp\!\left(\frac{C}{\sqrt{W}}\right),
\end{equation}
one obtains
\begin{equation}
\ln\!\left[R(W)-R_\infty\right]
=
\ln A
+
CW^{-1/2},
\end{equation}
which predicts that the experimental data should become linear when plotted against $W^{-1/2}$. Figure~\ref{fig:linearized} shows that this expectation is well satisfied by the digitized data extracted from Ref.~\cite{Yang2025PRL}. A linear regression yields a coefficient of determination $R^2=0.947$, supporting the reciprocal-space confinement scaling of the Zaccone theory \cite{Zaccone2025PRM}. For comparison, fitting the same digitized dataset with a conventional exponential dependence of the form $R(W)=R_\infty+A\exp(-BW)$ gives a slightly lower coefficient of determination, $R^2=0.927$. Overall, this independent reanalysis provides additional evidence that the thickness dependence of the Yang \textit{et al.} \cite{Yang2025PRL} data is consistent with the reciprocal-space confinement mechanism proposed in Ref.~\cite{Zaccone2025PRM}.

\subsection{Carrier density in ultra-thin films}

An important experimental question concerns whether the increase of resistivity in ultra-thin films originates solely from a reduction of the carrier mobility or whether the free-carrier concentration itself also becomes thickness dependent. Classical size-effect theories, including the Fuchs--Sondheimer and Mayadas--Shatzkes models, implicitly assume that the carrier concentration remains equal to its bulk value and that the entire thickness dependence of the conductivity arises from additional momentum-relaxing scattering processes.

A growing body of experimental evidence indicates that this assumption is not always valid in the nanometer regime. Hall-effect measurements by Du \textit{et al.}~\cite{Du2004} on ultrathin Al films revealed a systematic decrease of the apparent carrier concentration with decreasing film thickness, providing direct evidence that the free-electron population itself is modified by confinement. Similar departures from the bulk free-electron picture have also been discussed in nanostructured metallic films, where localization and confinement effects progressively modify the electronic response of the material \cite{Mirigliano2021,Mirigliano2020}.

Beyond electrical resistivity, the reciprocal-space confinement mechanism should also influence thermoelectric transport. Since the Seebeck coefficient is directly related to the energy dependence of the electronic conductivity, simultaneous measurements of resistivity and absolute thermoelectric power~\cite{Gasser2008} may provide a particularly sensitive experimental probe of confinement-induced modifications of the electronic density of states.

The electrical conductivity should therefore be written in its most general form as
\begin{equation}
\sigma(L)=e\,n(L)\mu(L),
\label{eq:sigma_Hall}
\end{equation}
or equivalently,
\begin{equation}
\rho(L)^{-1}=e\,n(L)\mu(L).
\end{equation}
A pronounced increase of resistivity may therefore originate from a reduction of the mobility, a reduction of the carrier concentration, or a combination of both effects. Distinguishing these two contributions experimentally requires simultaneous measurements of electrical resistivity and Hall carrier density.

Within the reciprocal-space confinement theory reviewed here, the thickness dependence of $n(L)$ arises naturally from the progressive suppression of available electronic states under strong spatial confinement. In semiconductors this mechanism leads directly to carrier depletion through the confinement-induced shift of the chemical potential, whereas in metals it reflects a reduction of the transport-active electronic states associated with the reconstruction of the Fermi surface. The emerging experimental evidence therefore suggests that thickness-dependent carrier concentration is an intrinsic consequence of quantum confinement rather than an incidental materials effect, highlighting the need for transport theories that go beyond mobility-based descriptions alone.

\subsection{Indirect evidence from superconductivity}
An important indirect validation of the reciprocal-space confinement model reviewed here is provided by recent applications to superconductivity. The same confinement-induced restructuring of the electronic states was shown to quantitatively predict the thickness dependence of the superconducting critical temperature $T_c$ in ultra-thin metallic films without adjustable parameters~\cite{Ummarino2025}. More recently, the theory was generalized within the Eliashberg theory framework to predict confinement-induced superconductivity in ultrathin noble-metal films (Au, Ag and Cu)~\cite{PhysRevMaterials.8.L101801}, a prediction that has recently received independent support from parameter-free \textit{ab initio} calculations~\cite{Zhang}. The ability of the same reciprocal-space confinement model to quantitatively account for both normal-state transport and superconducting properties strongly suggests that it captures a fundamental aspect of the electronic structure of ultra-thin metallic systems.

\section{Implications for nanoelectronics and interconnects}

The scaling of metallic interconnects has traditionally been limited by the increase of resistivity caused by surface and grain-boundary scattering. However, Eq.~\eqref{eq:FS_Zaccone} suggests that, below a few nanometers, a more severe limitation appears: the carrier population itself may be reduced by confinement. This implies that even ideal single-crystalline films with atomically smooth surfaces are not immune to resistivity enhancement.

The design of future interconnects and quantum devices must therefore account for three regimes:
(i) a bulk regime, $L\gg \ell$, where $\rho\approx\rho_0$;
(ii) a classical size-effect regime, $L\sim\ell$, where surface and grain-boundary scattering dominate; (iii) a quantum-confinement regime, $L\lesssim 10~{\rm nm}$, where the available electronic states and carrier density are modified. The third regime is especially important for atomically thin metals and next-generation interconnects~\cite{Kim2024Science,KhanPop2025}, quantum point contacts and atomic-scale conductors~\cite{Agrait2003}, memristive and neuromorphic devices~\cite{Milano2022,MemristiveReview2023}, nanoscale electrodes and electrical contacts in low-dimensional materials~\cite{ACSNanoContacts2025}, as well as ultra-scaled semiconductor architectures approaching the ultimate limits of Moore's law~\cite{HennessyPatterson2019,Shalf2020}.

\section{Outlook}

The emerging picture presented in this review suggests several promising directions for future research. From the experimental viewpoint, systematic Hall-effect measurements on ultra-thin metallic films will be essential for disentangling the respective roles of carrier density and mobility in the confinement regime. While the measurements of Yang \textit{et al.} \cite{Yang2025PRL} provide compelling evidence for the predicted exponential increase of resistivity, the corresponding evolution of the transport-active carrier concentration has not yet been determined experimentally. Equally important will be the use of epitaxial single-crystalline films, where grain-boundary scattering is largely eliminated and quantum confinement can be investigated in isolation. Extending the analysis of existing resistivity-versus-thickness datasets for Au, Ag, Cu, Al, Pt and technologically relevant semiconductors using the combined expression
\begin{equation}
\rho(L)=\rho_{\rm FS}(L)\exp\!\left(\frac{C}{\sqrt{L}}\right)
\end{equation}
may help establish the range of validity and universality of the confinement law across different classes of materials. An important experimental challenge in future work will be to perform Hall-effect measurements on epitaxial metallic nanofilms in the few-nanometer regime. 

From the theoretical perspective, the most important challenge is to derive the confinement factor $\eta_{\rm QC}(L)$ directly from the confined electronic structure of metallic films, thereby placing the metallic extension proposed here on the same microscopic footing as the analytical theory developed for semiconductors. Such a derivation would establish a direct quantitative connection between reciprocal-space topology and electrical transport without relying on phenomenological assumptions.

An equally promising direction is the integration of reciprocal-space confinement theory with parameter-free first-principles transport methods. Recent advances in atomistic descriptions of electron--surface scattering and quantum-confined metallic conductors provide an ideal platform for incorporating confinement-induced modifications of the electronic-state manifold into predictive transport calculations~\cite{Yuanyue,KhanPop2025,Valentinis2016}. Establishing this connection would unify semiclassical transport theory, electronic-structure calculations and quantum confinement within a common theoretical framework.

Finally, the concepts discussed here are likely to extend well beyond continuous metallic thin films. Similar confinement-induced modifications of the electronic state space are expected to influence transport in nanogranular and cluster-assembled metallic materials~\cite{Mirigliano2021}, atomically thin interconnects \cite{Empante2019}, quantum point contacts \cite{Beenakker}, memristive and neuromorphic devices \cite{Minnai2017,Minnai2018}, as well as superconducting ultra-thin films, where the same reciprocal-space confinement framework has recently been shown to quantitatively describe the thickness dependence of the superconducting critical temperature and to predict confinement-induced superconductivity in noble metals~\cite{Travaglino_2023,PhysRevMaterials.8.L101801,Zhang}. More generally, these developments suggest that reciprocal-space confinement may represent a unifying principle governing the electronic properties of matter under extreme spatial confinement. Developing a comprehensive microscopic theory capable of describing transport, superconductivity and other quantum phenomena within a single reciprocal-space framework constitutes one of the most exciting challenges for future research.

\section{Conclusions}

The progressive miniaturization of electronic devices has brought metallic and semiconducting films into a regime where the characteristic dimensions become comparable with fundamental electronic length scales. In this limit, the conventional picture of electrical transport based exclusively on scattering processes is no longer sufficient. While classical theories such as those of Fuchs--Sondheimer and Mayadas--Shatzkes successfully describe the mesoscopic regime, they do not account for the profound modification of the electronic state space that occurs under extreme spatial confinement.

The central message emerging from this review is that ultra-thin films exhibit a hierarchy of transport mechanisms. At large thicknesses, transport is governed by bulk scattering. As the film thickness approaches the electronic mean free path, surface and grain-boundary scattering progressively reduce the carrier mobility. Finally, in the few-nanometer regime, quantum confinement reconstructs the topology of the available electronic states in reciprocal space, suppressing long-wavelength states and fundamentally altering the phase space available for transport. This mechanism naturally gives rise to the characteristic exponential dependence
\begin{equation}
\rho(L)\sim\exp\!\left(\frac{C}{\sqrt{L}}\right),
\end{equation}
which has now been observed in both weakly doped semiconductor films~\cite{Duffy2019,Zaccone2025PRM} and, more recently, in single-crystalline Au nanofilms~\cite{Yang2025PRL}.

Taken together, these theoretical and experimental developments strongly suggest that reciprocal-space confinement is not merely a correction to classical size-effect theories, but rather represents a distinct transport mechanism that becomes dominant once the film thickness is reduced to only a few nanometers. While the exponential confinement law follows directly from the analytical theory for weakly doped semiconductors, the recent measurements on single-crystalline Au films provide compelling experimental evidence that the same reciprocal-space mechanism also extends to metallic systems. The proposed unified transport framework,
\begin{equation}
\sigma(L)=e\,n(L)\,\mu_{\rm FS}(L)\,\eta_{\rm QC}(L),
\end{equation}
provides a common language in which classical scattering, carrier depletion, and confinement-induced reduction of the transport-active reciprocal-space manifold appear as complementary aspects of the same microscopic description. In this sense, the present framework unifies the transport physics of semiconductors and metals, while naturally recovering the established classical theories as limiting cases.

Looking forward, an important challenge will be the development of a fully microscopic theory of the confinement factor $\eta_{\rm QC}(L)$ directly from the confined electronic structure, allowing quantitative predictions without phenomenological assumptions. Extending the present framework to multiband materials, strongly correlated systems, topological semimetals, two-dimensional conductors, and superconducting ultra-thin films also represents an exciting direction for future research. Such developments would establish reciprocal-space confinement as a general organizing principle governing the electronic properties of matter under extreme spatial confinement.

More broadly, the emergence of the same confinement physics in normal-state transport, carrier depletion, and superconductivity suggests that reciprocal-space topology may constitute one of the fundamental concepts required to understand the electronic behavior of low-dimensional materials. As electronic technologies continue to approach the atomic limit, incorporating confinement-induced modifications of the electronic state space into transport theory will likely become not only desirable but essential for the predictive design of future nanoelectronic and quantum devices.

\bibliographystyle{tfnlm}
\bibliography{refs}

@article{Fuchs1938,
  author  = {Fuchs, K.},
  title   = {The conductivity of thin metallic films according to the electron theory of metals},
  journal = {Mathematical Proceedings of the Cambridge Philosophical Society},
  volume  = {34},
  number  = {1},
  pages   = {100--108},
  year    = {1938},
  doi     = {10.1017/S0305004100019952}
}

@book{AshcroftMermin,
  author    = {Ashcroft, Neil W. and Mermin, N. David},
  title     = {Solid State Physics},
  publisher = {Holt, Rinehart and Winston},
  address   = {New York},
  year      = {1976},
  isbn      = {9780030839931}
}

@book{MottJones1936,
  author    = {Mott, N. F. and Jones, H.},
  title     = {The Theory of the Properties of Metals and Alloys},
  publisher = {Oxford University Press},
  address   = {Oxford},
  year      = {1936}
}

@book{Ziman1960,
  author    = {Ziman, J. M.},
  title     = {Electrons and Phonons: The Theory of Transport Phenomena in Solids},
  publisher = {Oxford University Press},
  address   = {Oxford},
  year      = {1960},
  isbn      = {9780198507796}
}

@article{Sondheimer1952,
  author  = {Sondheimer, E. H.},
  title   = {The mean free path of electrons in metals},
  journal = {Advances in Physics},
  volume  = {1},
  number  = {1},
  pages   = {1--42},
  year    = {1952},
  doi     = {10.1080/00018735200101151}
}

@article{Kim2024Science,
  author  = {Kim, J. S. and Gall, D. and Pop, Eric},
  title   = {Addressing Interconnect Challenges for Enhanced Microelectronics},
  journal = {Science},
  volume  = {386},
  number  = {6725},
  pages   = {eadk6189},
  year    = {2024},
  doi     = {10.1126/science.adk6189}
}

@article{ACSNanoContacts2025,
  author  = {Zhang, Y. and Liu, Y.},
  title   = {Recent Contact Strategies for Two-Dimensional Electronics},
  journal = {ACS Nano},
  volume  = {19},
  number  = {1},
  year    = {2025},
  doi     = {10.1021/acsnano.5c07026}
}

@article{Agrait2003,
  author  = {Agraït, Nicol{\'a}s and Yeyati, Alfredo Levy and van Ruitenbeek, Jan M.},
  title   = {Quantum Properties of Atomic-Sized Conductors},
  journal = {Physics Reports},
  volume  = {377},
  number  = {2--3},
  pages   = {81--279},
  year    = {2003},
  doi     = {10.1016/S0370-1573(02)00633-6}
}

@article{Gasser2008,
  author  = {Jean-Georges Gasser},
  title   = {Understanding the resistivity and absolute thermoelectric power of disordered metals and alloys},
  journal = {Journal of Physics: Condensed Matter},
  volume  = {20},
  number  = {11},
  pages   = {114230},
  year    = {2008},
  doi     = {10.1088/0953-8984/20/11/114230}
}

@article{MayadasShatzkes1970,
  author  = {Mayadas, A. F. and Shatzkes, M.},
  title   = {Electrical-resistivity model for polycrystalline films: the case of arbitrary reflection at external surfaces},
  journal = {Physical Review B},
  volume  = {1},
  number  = {4},
  pages   = {1382--1389},
  year    = {1970},
  doi     = {10.1103/PhysRevB.1.1382}
}

@article{Zhang2004,
  author  = {Zhang, W. and Brongersma, S. H. and Richard, O. and Brijs, B. and Palmans, R. and Froyen, L. and Maex, K.},
  title   = {Influence of the electron mean free path on the resistivity of thin metal films},
  journal = {Microelectronic Engineering},
  volume  = {76},
  number  = {1--4},
  pages   = {146--152},
  year    = {2004},
  doi     = {10.1016/j.mee.2004.07.041}
}

@article{Du2004,
  author  = {Du, Hao and Gong, Jun and Sun, Chao and Lee, Soo Wohn and Wen, Li Shi},
  title   = {Carrier density and DC conductivity of ultrathin aluminum films},
  journal = {Journal of Materials Science},
  volume  = {39},
  pages   = {2865--2867},
  year    = {2004},
  doi     = {10.1023/B:JMSC.0000021487.40207.60}
}

@article{Duffy2019,
  author  = {MacHale, John and Meaney, Fintan and Kennedy, Noel and Eaton, Luke and Mirabelli, Gioele and White, Mary and Thomas, Kevin and Pelucchi, Emanuele and Petersen, Dirch Hjorth and Lin, Rong and Petkov, Nikolay and Connolly, James and Hatem, Chris and Gity, Farzan and Ansari, Lida and Long, Brenda and Duffy, Ray},
  title   = {Exploring conductivity in ex-situ doped Si thin films as thickness approaches 5 nm},
  journal = {Journal of Applied Physics},
  volume  = {125},
  number  = {22},
  pages   = {225709},
  year    = {2019},
  doi     = {10.1063/1.5098307}
}

@article{Zaccone2025PRM,
  author  = {Zaccone, Alessio},
  title   = {Thickness-dependent conductivity of nanometric semiconductor thin films},
  journal = {Physical Review Materials},
  volume  = {9},
  pages   = {046001},
  year    = {2025},
  doi     = {10.1103/PhysRevMaterials.9.046001}
}

@article{Yang2025PRL,
  author  = {Yang, Xiaomeng and Zhang, Jianfei and Li, Wei and Zhang, Qing and Zhou, Quan and Mao, Shengcheng and Wang, Menglong and Zheng, Sikang and Wang, Lihua and Zhang, Ze and Han, Xiaodong},
  title   = {Approaching the Ballistic Transport Limit in Single Crystalline {Au}},
  journal = {Physical Review Letters},
  volume  = {135},
  pages   = {216302},
  year    = {2025},
  doi     = {10.1103/h82y-wds1}
}

@article{KhanPop2025,
  author  = {Khan, Asir Intisar and Ramdas, Akash and Lindgren, Emily and Kim, Hyun-Mi and Won, Byoungjun and Wu, Xiangjin and Saraswat, Krishna and Chen, Ching-Tzu and Suzuki, Yuri and da Jornada, Felipe H. and Oh, Il-Kwon and Pop, Eric},
  title   = {Surface conduction and reduced electrical resistivity in ultrathin noncrystalline NbP semimetal},
  journal = {Science},
  volume  = {387},
  number  = {6729},
  pages   = {62--67},
  year    = {2025},
  doi     = {10.1126/science.adq7096}
}

@article{Valentinis2016,
  author  = {Valentinis, D. and van der Marel, D. and Berthod, C.},
  title   = {Rise and fall of shape resonances in thin films of BCS superconductors},
  journal = {Physical Review B},
  volume  = {94},
  number  = {5},
  pages   = {054516},
  year    = {2016},
  doi     = {10.1103/PhysRevB.94.054516}
}

@article{Moore1965,
  author  = {Moore, Gordon E.},
  title   = {Cramming More Components onto Integrated Circuits},
  journal = {Electronics},
  volume  = {38},
  number  = {8},
  pages   = {114--117},
  year    = {1965}
}

@article{Dennard1974,
  author  = {Dennard, Robert H. and Gaensslen, Fritz H. and Yu, Hwa-Nien and Rideout, V. Leo and Bassous, Ernest and LeBlanc, Andre R.},
  title   = {Design of Ion-Implanted MOSFET's with Very Small Physical Dimensions},
  journal = {IEEE Journal of Solid-State Circuits},
  volume  = {9},
  number  = {5},
  pages   = {256--268},
  year    = {1974},
  doi     = {10.1109/JSSC.1974.1050511}
}

@article{HennessyPatterson2019,
  author  = {Hennessy, John L. and Patterson, David A.},
  title   = {A New Golden Age for Computer Architecture},
  journal = {Communications of the ACM},
  volume  = {62},
  number  = {2},
  pages   = {48--60},
  year    = {2019},
  doi     = {10.1145/3282307}
}

@article{Shalf2020,
  author  = {Shalf, John},
  title   = {The Future of Computing Beyond Moore's Law},
  journal = {Philosophical Transactions of the Royal Society A},
  volume  = {378},
  number  = {2166},
  pages   = {20190061},
  year    = {2020},
  doi     = {10.1098/rsta.2019.0061}
}

@article{Davies2025,
  author  = {Davies, Michael},
  title   = {Envisioning the Economics of a Semiconductor Revolution},
  journal = {Communications of the ACM},
  volume  = {68},
  number  = {5},
  pages   = {44--49},
  year    = {2025},
  doi     = {10.1145/3711920}
}

@article{MemristiveReview2023,
  author  = {Xia, Qiangfei and Yang, J. Joshua},
  title   = {Recent Advances and Future Prospects for Memristive Devices},
  journal = {ACS Nano},
  volume   = {17},
  number   = {14},
  pages    = {13254--13286},
  year     = {2023},
  doi      = {10.1021/acsnano.3c03505}
}

@article{Milano2022,
  author  = {Milano, Gianluca and Pedretti, Giacomo and Montesi, Lorenzo and Ricci, Stefano and Boarino, Luca and Pirri, Candido F. and Ielmini, Daniele},
  title   = {Quantum Conductance in Memristive Devices: Fundamentals, Developments, and Applications},
  journal = {Advanced Materials},
  volume   = {34},
  number   = {32},
  pages    = {2201248},
  year     = {2022},
  doi      = {10.1002/adma.202201248}
}

@article{Beenakker,
  author  = {B. J. van Wees and H. van Houten and C. W. J. Beenakker and J. G. Williamson and L. P. Kouwenhoven and D. van der Marel and C. T. Foxon},
  title   = {Quantized Conductance of Point Contacts in a Two-Dimensional Electron Gas},
  journal = {Physical Review Letters},
  volume  = {60},
  number  = {9},
  pages   = {848--850},
  year    = {1988},
  doi     = {10.1103/PhysRevLett.60.848}
}

@article{Empante2019,
  author  = {Thomas A. Empante and Aimee Martinez and Michelle Wurch and Yanbing Zhu and Adane K. Geremew and Koichi Yamaguchi and Miguel Isarraraz and Sergey Rumyantsev and Evan J. Reed and Alexander A. Balandin and Ludwig Bartels},
  title   = {Low Resistivity and High Breakdown Current Density of 10-nm Diameter van der Waals TaSe$_3$ Nanowires by Chemical Vapor Deposition},
  journal = {Nano Letters},
  volume  = {19},
  number  = {7},
  pages   = {4355--4361},
  year    = {2019},
  doi     = {10.1021/acs.nanolett.9b00958}
}

@misc{Zhang,
      title={Enhanced superconductivity in atomically thin noble metals: From quantum confinement to interface-induced Lifshitz transition}, 
      author={Chun-Jie Zhang and Bing Zhang and Yapeng Wu and Xiao-Ping Li and Lei Wang},
      year={2026},
      eprint={2606.03663},
      archivePrefix={arXiv},
      primaryClass={cond-mat.supr-con},
      url={https://arxiv.org/abs/2606.03663}, 
}

@article{Ummarino2025,
doi = {10.1088/1361-648X/ad92ed},
url = {https://doi.org/10.1088/1361-648X/ad92ed},
year = {2024},
month = {nov},
publisher = {IOP Publishing},
volume = {37},
number = {6},
pages = {065703},
author = {Ummarino, Giovanni Alberto and Zaccone, Alessio},
title = {Quantitative {Eliashberg} theory of the superconductivity of thin films},
journal = {Journal of Physics: Condensed Matter}
}

@article{PhysRevMaterials.8.L101801,
  title = {Can the noble metals (Au, Ag, and Cu) be superconductors?},
  author = {Ummarino, Giovanni Alberto and Zaccone, Alessio},
  journal = {Phys. Rev. Mater.},
  volume = {8},
  issue = {10},
  pages = {L101801},
  numpages = {6},
  year = {2024},
  month = {Oct},
  publisher = {American Physical Society},
  doi = {10.1103/PhysRevMaterials.8.L101801},
  url = {https://link.aps.org/doi/10.1103/PhysRevMaterials.8.L101801}
}

@article{Minnai2017,
  author  = {Chlo{\'e} Minnai and Andrea Bellacicca and Simon A. Brown and Paolo Milani},
  title   = {Facile Fabrication of Complex Networks of Memristive Devices},
  journal = {Scientific Reports},
  volume  = {7},
  pages   = {7955},
  year    = {2017},
  doi     = {10.1038/s41598-017-08244-y}
}

@article{Minnai2018,
  author  = {Chlo{\'e} Minnai and Matteo Mirigliano and Simon A. Brown and Paolo Milani},
  title   = {The Nanocoherer: An Electrically and Mechanically Resettable Resistive Switching Device Based on Gold Clusters Assembled on Paper},
  journal = {Nano Futures},
  volume  = {2},
  number  = {1},
  pages   = {011002},
  year    = {2018},
  doi     = {10.1088/2399-1984/aab4ee}
}

@book{Kittel,
title = {Introduction to solid state physics},
author = {Kittel, C.},
publisher = {John Wiley and Sons},
place = {Hoboken, NJ},
year = {2005}
}

@article{Heat,
  title = {Quantum confinement theory of the heat capacity of thin films},
  author = {Zaccone, Alessio},
  journal = {Phys. Rev. Mater.},
  volume = {8},
  issue = {5},
  pages = {056001},
  numpages = {5},
  year = {2024},
  month = {May},
  publisher = {American Physical Society},
  doi = {10.1103/PhysRevMaterials.8.056001},
  url = {https://link.aps.org/doi/10.1103/PhysRevMaterials.8.056001}
}

@article{Mirigliano2021,
  author  = {Mirigliano, Matteo and Milani, Paolo},
  title   = {Electrical Conduction in Nanogranular Cluster-Assembled Metallic Films},
  journal = {Advances in Physics: X},
  volume  = {6},
  number  = {1},
  pages   = {1908847},
  year    = {2021},
  doi     = {10.1080/23746149.2021.1908847}
}

@article{Mirigliano2020,
  author  = {Mirigliano, Matteo and Radice, Stefano and Falqui, Andrea and Casu, Alberto and Cavaliere, Fabio and Milani, Paolo},
  title   = {Anomalous Electrical Conduction and Negative Temperature Coefficient of Resistance in Nanostructured Gold Resistive Switching Films},
  journal = {Scientific Reports},
  volume  = {10},
  pages   = {20482},
  year    = {2020},
  doi     = {10.1038/s41598-020-76632-y}
}

@article{thermal,
    author = {Zaccone, Alessio},
    title = {Phonon-confinement theory of thermal conductivity in ultrathin silicon films},
    journal = {Journal of Applied Physics},
    volume = {138},
    number = {22},
    pages = {225305},
    year = {2025},
    month = {12},
    issn = {0021-8979},
    doi = {10.1063/5.0304896},
    url = {https://doi.org/10.1063/5.0304896},
    eprint = {https://pubs.aip.org/aip/jap/article-pdf/doi/10.1063/5.0304896/20835379/225305_1_5.0304896.pdf},
}

@article{ZhouGall2018,
  author = {Zhou, Tianji and Gall, Daniel},
  title = {Resistivity Scaling Due to Electron Surface Scattering in Thin Metal Layers},
  journal = {Physical Review B},
  volume = {97},
  pages = {165406},
  year = {2018},
  doi = {10.1103/PhysRevB.97.165406}
}

@article{Yuanyue,
author = {Zhang, Chenmu and Liu, Yuanyue},
title = {Electron-Surface Scattering from First-Principles},
journal = {ACS Nano},
volume = {18},
number = {40},
pages = {27433-27439},
year = {2024},
doi = {10.1021/acsnano.4c07698},
    note ={PMID: 39325662},

URL = { 
    
        https://doi.org/10.1021/acsnano.4c07698
    
    

},
eprint = { 
    
        https://doi.org/10.1021/acsnano.4c07698
    
    

}

}

@article{Travaglino_2023,
    author = {Travaglino, Riccardo and Zaccone, Alessio},
    title = "{Extended analytical BCS theory of superconductivity in thin films}",
    journal = {Journal of Applied Physics},
    volume = {133},
    number = {3},
    pages = {033901},
    year = {2023},
    month = {01},
    issn = {0021-8979},
    doi = {10.1063/5.0132820},
    url = {https://doi.org/10.1063/5.0132820},
    eprint = {https://pubs.aip.org/aip/jap/article-pdf/doi/10.1063/5.0132820/16766705/033901\_1\_online.pdf},
}

@article{Travaglino_2022,
doi = {10.1088/1361-6455/ac5583},
url = {https://dx.doi.org/10.1088/1361-6455/ac5583},
year = {2022},
month = {mar},
publisher = {IOP Publishing},
volume = {55},
number = {5},
pages = {055301},
author = {Riccardo Travaglino and Alessio Zaccone},
title = {Analytical theory of enhanced Bose–Einstein condensation in thin films},
journal = {Journal of Physics B: Atomic, Molecular and Optical Physics}
}

\end{document}